This is the author's accepted manuscript of the article published in *Physical Review E* 105, 044701 (2022). The official Version of Record is available online at https://doi.org/10.1103/PhysRevE.105.044701. This work was supported by the National Science Centre, Poland, under grant No. 2018/31/B/ST3/03609 and project No. 2020/39/O/ST5/03460.

# Structure of the twist-bend nematic phase with respect to the orientational molecular order of the thioether-linked dimers

*Antoni Kocot,[1‡] Barbara Loska,[1] Yuki Arakawa,[2] Katarzyna Merkel[1‡]**

[1]Institute of Materials Engineering, Faculty of Science and Technology, University of Silesia, ul. 75. Pułku Piechoty, Chorzów 41-500

[2] Department of Applied Chemistry and Life Science, Graduate School of Engineering, Toyohashi University of Technology, Toyohashi, 441-8580, Japan

KEYWORDS: Orientational order, biaxiality, liquid crystal dimers, twist-bend phase, thioether dimers

ABSTRACT

An analysis of the IR absorbance for the segmented functional groups of liquid crystal dimers: mesogen and linker, enabled the orientation order to be determined and information about the dipole interactions in the nematic and twist-bend nematic phases to be obtained. The long axis orientational order increases as the temperature decreases in the nematic phase, although much more slowly than for the classical nematics, and then reverses this trend in the twist-bend nematic phase due to the tilt of the molecules. In the nematic phase, the short axis of the molecule performs an isotropic uniform rotation and has a uniaxial alignment. In the twist-bend nematic phase, however, biaxial ordering occurs and grows significantly in accordance with the helical deformation of the director. Changes in the mean absorbance in the twist-bend nematic phase were observed: a decrease for the longitudinal dipole at the nematic-twist-bend nematic phase transition, thus emphasizing the antiparallel axial interaction of the dipoles, while the absorbance of the transverse dipoles remains unchanged up to 340 K, and then the latter become parallelly correlated.

# I. INTRODUCTION

An ongoing topic in the science of liquid crystals (LC) is understanding the interrelationships between the molecular shape and macroscopic self-organization and creating new ordering systems with a new symmetry. The problem of symmetry breaking and symmetry transformation from a molecular to macroscopic (phase) system is one of the most important issues in searching for and studying new LC materials. A perfect example of such a situation is the discovery of the spontaneous formation of the chiral order in fluids, as was evidenced by the macroscopic chiral domains in liquids of achiral bent molecules [1–2]. The latest and most exciting manifestation of spontaneous chirality in the LC of achiral molecules is the twist-bend nematic ($N_{TB}$) phase, which was initially proposed theoretically [3–6], then experimentally confirmed [7–11] and intensively studied in many systems of bent molecules [12–25], in particular, the cyanobiphenyl (CB) dimers (CBCnCB, n = 7.9,11) [8,9,11,15,26–28] and terphenyl-based dimers (DTCnC5, n = 7,9,11) [29–33]. In the $N_{TB}$ phase, the molecules simultaneously bend and twist in space forming a helix of the average long molecular axis (director). The structure of the $N_{TB}$ phase lacks a coherent modulation of the density that accompanies the helix [34], and resonant X-ray diffraction, which describes the molecular orientation, revealed a helical phase periodicity [20,24,27,28,35–41]. An unusual feature of this phase is its very short pitch, which is often much smaller than 10 nm. How the helical phase remains a fully three-dimensional liquid in the presence of such a strong, consistent internal orientation order remains an open question. It is also puzzling why molecules with a curved shape first form a nematic (N) phase, which is characteristic of calamitic molecules, and then a twist-bend phase? In this paper, we will try to answer the above-mentioned questions. We employed Fourier transform infrared spectroscopy (FTIR) to study the orientational order of cyanobiphenyl-based dimer series to solve this problem. Infrared spectroscopy for anisotropic systems provides information on the orientation of the individual functional groups of molecules as well as specific intra- and intermolecular interactions. The single most characteristic property of a liquid crystalline phase is the orientational order of its constituent molecules as measured by second-rank order parameters, $S_{\alpha\beta}$, which was introduced by Saupe for the uniaxial phases: [42,43]

$$S^{l}_{\alpha\beta} = \langle \tfrac{1}{2}\left(3 l_{i,\alpha} l_{i,\beta} - \delta_{\alpha\beta}\right)\rangle \tag{1}$$

Where: 〈 〉 signifies the average and $l_{i,\alpha}$, $l_{i,\beta}$ are the cosine of the angle between the molecular axis $\alpha$,$\beta$ and the laboratory axis $i$ $(i=X,Y,Z)$.

Determining the degree of ordering becomes possible if individual components of absorbance in a laboratory system are expressed as a function of the orientation of the corresponding transition dipole moments as below:

$$A_{X,Y}/A_0 = 1 + S\left(\frac{3}{2}\sin^2\sigma - 1\right) + \frac{1}{2}D\sin^2\sigma\cos 2\varphi \tag{2a}$$

$$A_Z/A_0 = 1 + S(2 - 3\sin^2\sigma) - D\sin^2\sigma\cos 2\varphi \tag{2b}$$

Where: $A_0$ is the average absorbance equal to $(A_X+A_Y+A_Z)/3$, $\sigma$ is the polar angle between the transition dipole ($\mu$) and the ***z***-axis of the molecule and $\varphi$ is the azimuthal angle that the transition dipole makes with the ***x-z*** plane in a molecular system (see Figure 1). The various values of the $\sigma$ angle for the different bands were obtained from a molecular structure simulation (see paragraph "RESULTS AND DISCUSSION"). In the paper for bent dimers, the temperature dependencies of two orientational order parameters were determined: one that describes the orientational order of the long axis ($S$-parameter) and one that describes the molecular biaxiality parameter $D$:

$$S = S^{Z}_{zz}, D = S^{Z}_{xx} - S^{Z}_{yy} \tag{3}$$

Parameter $S$ is the measure of the increase in the compatibility of the long molecular ***z***-axis with the ***Z***-axis of the laboratory reference system. At the same time, $D$ describes the rotational biasing of the short molecular axes.

Here, we report on an FTIR study for two groups of dimers: a symmetric one containing the thioether linking groups (C-S-C) with the acronym CBSCnSCB (n=5,7) and the ether linking groups (C-O-C) with the abbreviation CBOCnOCB (n=7) and an asymmetric one that contains both the ether and thioether linking groups, CBSCnOCB (n=5,7). By comparing the ordering parameters for different dimer segments: the cyanobiphenyl core, which represents the dimer arms, the central alkyl linker and the bridges connecting the rigid core to the linker, we investigated the bending of the molecule in the N phase. In the $N_{TB}$ phase, the orientational order parameter reverses its trend due to the helical tilt of the director. Using the $N_{TB}$ order parameter ratio and the extrapolated trend from the N phase, we calculated the tilt angle of the molecules in the $N_{TB}$ phase. The assessment of the local director bending might also be useful [44,45]. Based on the obtained results combined with the data that was obtained from the resonant X-ray scattering [24,41], can we answer the question of whether the molecular biaxiality parameter predicts the local director bend, and consequently, the periodicity of a helical structure?

## II. EXPERIMENTAL SECTION

### A. Materials.

The bending angle of the molecules and the stability of the $N_{TB}$ phase are largely dependent on the nature of the bonds connecting the rigid mesogenic arms and the linker. Two groups of dimers were investigated in the study: symmetrical and asymmetric dimers, the mesogenic group of which is cyanobiphenyl (CB), and the linker contained seven or nine units / functional groups (methylene, thioether or ether groups). The series of symmetrical dimers included the material with the acronym CBC9CB, which contained nine methylene groups in a linker, then the material CBOC7OCB, in which an ether bridge linked the cyanobiphenyls with an alkyl chain with seven methylene groups and the thioether dimers (CBSCnSCB), in which a thioether bridge linked the cyanobiphenyls with a chain alkyl with five or seven methylene groups. In the asymmetric dimers with the acronym CBSCnOCB, the mesogens were linked on one side to an alkyl chain by a thioether bridge and on the other by an ether bridge. All of the thioether/ether compounds (CBSCnSCB, CBSCnOCB, CBOC7OCB) were synthesized and developed in collaboration with Professor Yuki Arakawa from the Department of Organic Chemistry of the Toyohashi University of Technology in Japan [46–47].

### B. FTIR Experiment.

The samples for the IR studies were aligned between two optically polished zinc selenide (ZnSe) and potassium bromide (KBr) windows. The windows were spin-coated with an SE-130 commercial polymer aligning agent (Nissan Chemical Industries, Ltd) in order to obtain a homogeneous alignment. The cells were assembled with a parallel positioning of the rubbing direction, and mylar foil was used as a spacer. The thickness of the fabricated cells was determined to be in the range of 5.1-5.6 μm by the measurements on the interference fringes using a spectrometer interfaced with a PC (Avaspec-2048). The samples were capillary filled by heating an empty cell five degrees below the transition to the isotropic phase in the N phase. The quality of the alignment was tested using polarizing microscopy. The textures of the samples were monitored using a polarizing microscope (Olympus BX56). Figure 1 shows the structure of a thioether dimer (CBSC7CB) as an example and the molecular and laboratory reference system that was used in the FTIR experiment and in the calculation of the order parameters.

In this study, an Agilent Cary 670 FTIR spectrometer was used to record the infrared spectra. The spectra for the ordered samples were recorded at a resolution of 2 $cm^{-1}$. In order to improve the data quality, the spectra were collected multiple times (32 scans). The measurements were taken using slow cooling and heating at a rate of 0.5 K/min. For the samples with a thioether bridge (CBSC7SCB, CBSC7OCB), an additional measurement was taken with a faster cooling rate (4 K/min) in a temperature range from a few degrees above the N-$N_{TB}$ transition until the solid phase (glass or crystal) was obtained. The faster cooling rate near the N-$N_{TB}$ transition temperature had a fundamental impact on the width of the $N_{TB}$ phase for the samples with a sulfide bridge. The temperature of the samples was stabilized using a PID temperature controller with an accuracy of 2 mK. The experiment was performed using the transmission method with a polarized IR beam. An IR-KRS5 grid polarizer was used to polarize the IR beam. The IR spectra were measured as a function of the polarizer rotation angle in the range 500-4000 $cm^{-1}$ of wavenumbers.

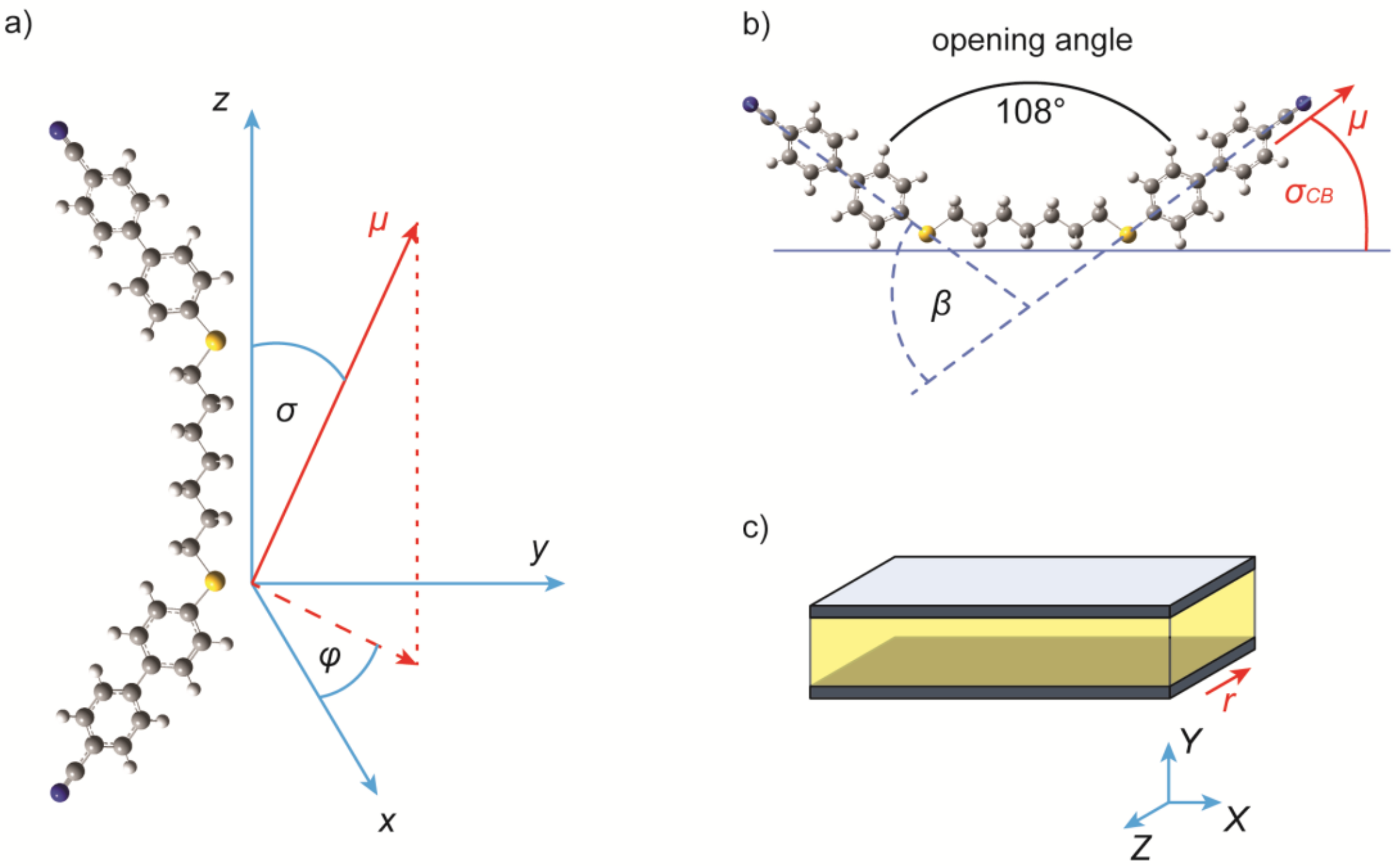

FIG 1. (a) Molecular Frame of Reference: ***z*** – long axis (bowstring), ***x***-axis normal to the bent plane, ***y*** – bow arrow axis, **σ** – polar angle (between transition dipole, **μ**, and the ***z***-axis of the molecule), **φ** is the azimuthal angle that the transition dipole makes with the ***x-z*** plane. (b) Optimized molecular structure of the CBSC7SCB molecule using the DFT method (B3LYP/G-311G(d,p)). $\sigma_{CB}$ – is the polar angle of the arm ($\sigma_{CB} = (180° - 108°)/2 = 36°$). (c) The orientation of the laboratory frame ($X,Y,Z$) for the planar sample**.** *In the nematic phase*, ***Z*** was an axis along and ***Y*** was perpendicular to the optical axis (the optical axis coincided with the rubbing direction). *In the $N_{TB}$ phase*, Z coincided with the helix axis., which was the symmetry elements of the $N_{TB}$ phase.

## III. RESULTS AND DISCUSSION

### A. Orientational order in the nematic phase

The analysis of the IR spectra towards the calculation orientational order requires knowledge of various angles in the molecular system. Some of these angles are acknowledged to be critically important for the calculations. Molecular structures of the CBSC5SCB, CBSC7SCB CBSC5OCB and CBSC7OCB molecules and the harmonic vibrational force constants, and absolute IR intensities were calculated using a density functional theory (DFT) (B3LYP/G-311G (d,p)). Some details are described in Supplementary Materials. For the study the structure in a realistic density of a system, we also optimise the molecule structure in a system with its nearest neighbouring molecules for symmetric dimers CBSC7SCB, see Figure S1 (*Supplementary Materials*). The Density-functional theory (DFT) calculations of the theoretical vibrational frequencies for symmetric dimers predict the coupling between vibrations in the dimer's same units (tween units). As a result, two vibrations are expected: the first due to the asynchronous vibration of the corresponding units and the second due to their synchronous swing [32,33,48]. Suppose the elementary units vibrate along the para axis of the rigid core (arm). In that case, the asynchronous vibration produces a transition dipole along the bowstring axis of the dimer. In contrast, the transition dipole has a synchronous swing, which is along the bow arrow axis. The frequencies of the asynchronous mode are higher than the synchronous frequencies. The coupling is significant if the vibrations involve the linker group, then the frequencies of the asynchronous mode are significantly higher than the synchronous frequencies. Otherwise, the coupling is relatively weak, and the resulting frequencies of the synchronous and asynchronous modes are very close to each other and, in reality, not possible to be deconvoluted. In practice, such vibrations can be treated as uncoupled. Thus, their dipole moment remains oriented almost along the para axis of the mesogen core unit.

The simplest way to calculate the *S* parameter is to use a band with a longitudinal transition dipole ($\mu_{||}$). Then, the *S*-order parameter can be calculated using the $A_Z$ or $A_X$ component of the absorbance, eq. (2). Please remember that $A_0$ is temperature dependent and is calculated as $A_0=(A_Z+2*A_X)/3$. It can be shown that for symmetric dimers (CBSC5SCB and CBSC7SCB), only the 1098 $cm^{-1}$ band has a transition dipole along the long molecular axis. Moreover, for the asymmetric dimers CBSC5OCB and CBSC7OCB, a few bands, e.g., 1098 $cm^{-1}$ and 1000 $cm^{-1}$, are almost along the long molecular axis. The band that is located at the wavenumber of 1098 $cm^{-1}$ can be assigned to the deformation vibrations of the C-H group in the benzene ring plane (*βCH ip CB, ip – in the plane of benzene* ). This vibration is complex and also involves a thioether linkage and therefore is associated with the asymmetric stretching vibration of the $C_{Ar}$-S group (*βCH ip CB* + $\nu_{as}C_{Ar}$-*S*, *Ar* – aromatic ring). This vibration is extremely sensitive to any conformational changes of the dimer. On the other hand, the band at 1000 $cm^{-1}$ is assigned to the breathing deformation vibration of the C-C atoms in the benzene ring (*βCC ip CB*). The polar angle can be estimated to be $\sigma< 5^o$. The comparison of the experimental spectrum of the CBS7SCB with the theoretical spectrum for different conformations is presented in Figures S2 and S3 (supplementary materials). For such a case, the second term in eq. (2a) can be neglected (as the *D* parameter in the nematic phase is expected to be small) and the order parameter can be calculated directly using the absorbances of a single band.

$$S = (A_Z/A_0 - 1)/(2 - 3\sin^2\sigma) \tag{4}$$

Figure 2 shows the order parameters of the long molecule axis for the asymmetric dimers (CBS5OCB, CBS7OCB), which were calculated from the 1000 $cm^{-1}$ band and for the symmetric dimers (CBS5SCB, CBS7SCB), which were calculated from the 1098 $cm^{-1}$ band.

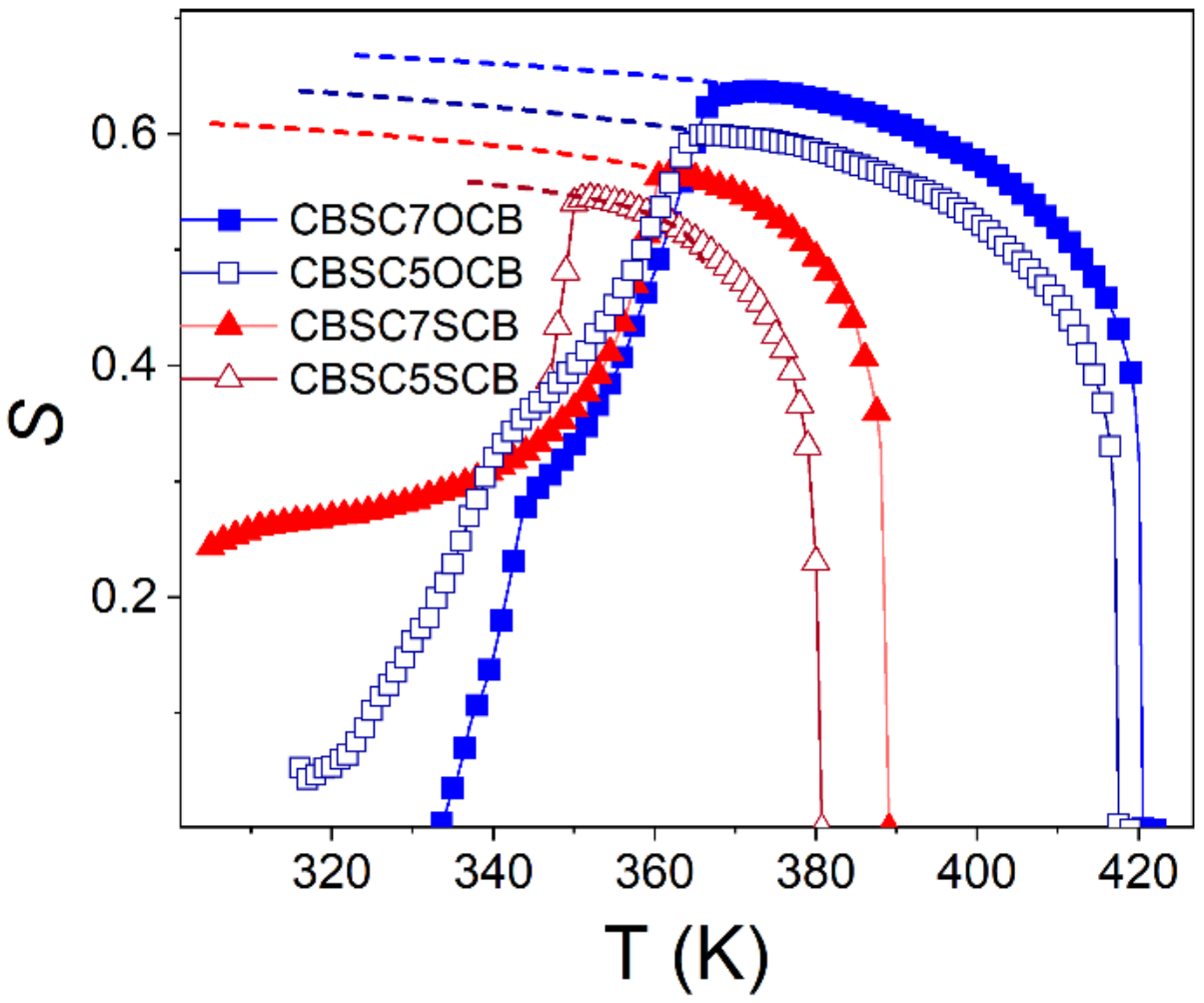

FIG. 2. Comparison of the *S*-order parameters of the long molecule axis of the thioether dimers. Calculated from the 1098 $cm^{-1}$ band for the symmetric dimers (CBSCnSCB) and from the1000 $cm^{-1}$ band for the asymmetric dimers (CBSCnOCB). *Dashed lines* – Fitting exp. data using the power law eq. (5). *Symmetric dimers:* △ – CBSC5SCB, ▲ – CBSC7SCB, *Asymmetric dimers:* □ – CBSC5OCB, ■ – CBSC7OCB.

However, for the other cyanobiphenyl bands (2200, 1600, 1485 $cm^{-1}$), the transition dipole was significantly inclined from the long molecular axis because the vibrations of the counterparts in the dimer were almost decoupled from one another (see. Fig. 3). The inclination angle was somehow associated with the bend angle $\sigma \cong \beta/2$, see Fig. 1b, but also depended on the population of other (not all-*trans*) conformers. Figure 3 shows the order parameters of the rigid core (para axis of the cyanobiphenyl), which was calculated from the 1600 cm$^{-1}$ band absorbances that can be assigned to the benzene ring vibration (*νCC ip CB*). It is clear that the more bent dimers were less compatible for creating the nematic order. Thus, the inclination angle must have a significant impact on the orientational order parameter. Increasing the inclination angle reduced orientational order relatively, i.e., the temperature dependence became weaker with a decrease in temperature. As a result, the whole range of *S* dependence in the N phase could not be reproduced by a power law (with the one critical exponent):

$$S = S_0\left(1 - \frac{T}{T_C}\right) \tag{5}$$

This behavior was more evident when we compared the temperature dependencies of the dimers that formed the $N_{TB}$ phase and had the same number of atoms in the linkage as that of the CBO7OCB dimer, see Figure 3. The temperature dependence of the $S_{CB}$ parameter for the CBO7OCB dimer, which was calculated for the 1600 cm$^{-1}$ band, could be well fitted by power law (5) with the critical exponent ≅0.20. Conversely, for the dimers that formed the $N_{TB}$ phase, the temperature dependence of $S_{CB}$ became significantly less steep after cooling.

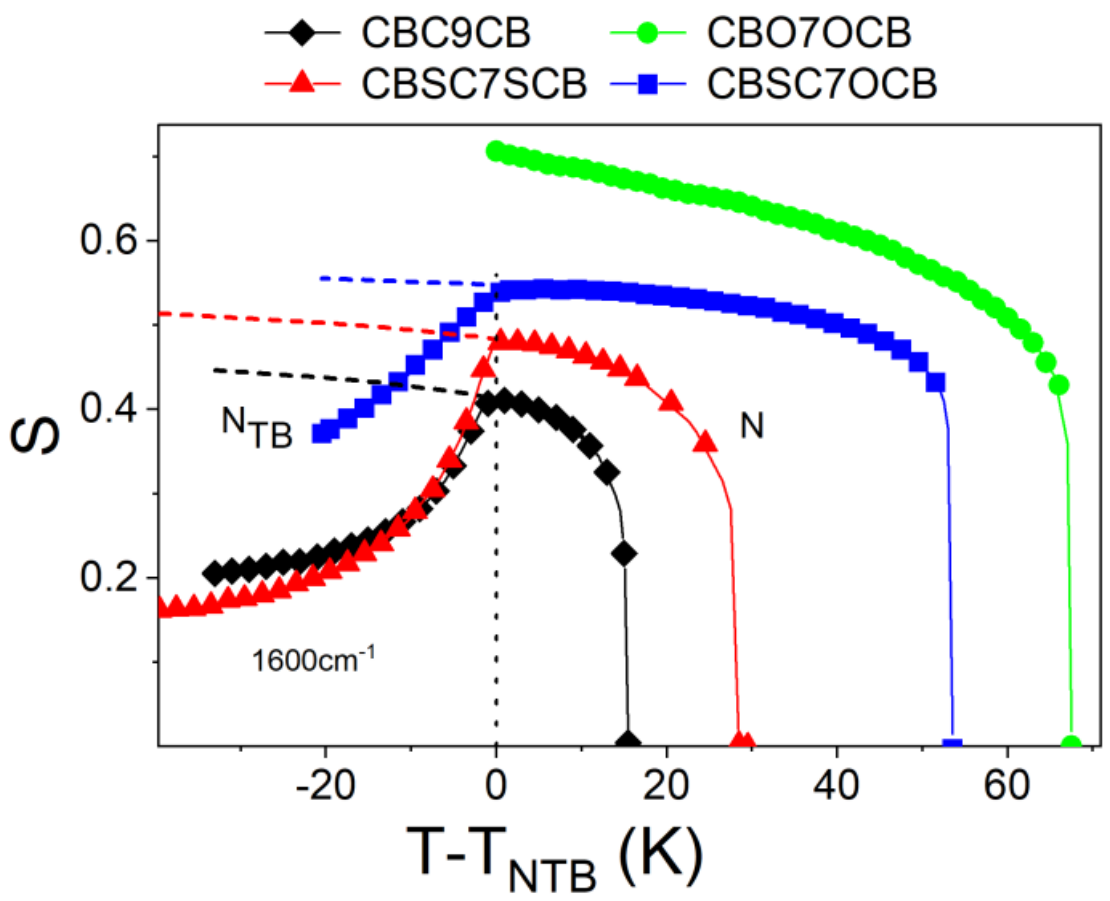

FIG. 3. Comparison of the $S_{CB}$-order parameters of the para axis of the rigid core (arms of the dimer). Calculated from the 1600 cm$^{-1}$ band for all of the dimers. *Symbols:* ● – CBOC7OCB, ■ – CBSC7OCB, ▲ – CBSC7SCB, ◆ – CBC9CB. *Dashed lines* – Fitting exp. data using the power law eq.(5).

It appeared that they formed a twist-bend deformation locally. This can be explained by the condensation of the structure, which was determined by the free volume and dense packing of the molecules, which are dependent on molecular shape [49,50]. It is crucial to estimate the inclination angle, i.e., the polar angle of the CB group $\sigma_{CB}$ as it can provide information about the behavior of the molecular bends with temperature. To do this, one should first calculate the order parameter, $S_{CB}$, of the para axis of the cyanobiphenyl unit, $S_{CB}$=($A_Z/A_0$-1)/2. The latter is understood as a *S*-parameter of the molecule long axis by a factor $3/2\cos^2\sigma_{CB}-1$, which is the Legendre polynomial $P_2$:

$$S_{CB} = S \cdot P_2(\cos\sigma_{CB}) \quad (6)$$

Then, we can estimate $\sigma_{CB}$, which is an argument of the $P_2$ Legendre polynomial. The inclination angle was found to be 23°, 20° 15° and 13° for CBSC5SCB, CBSC7SCB, CBSC5OCB and CBSC7OCB, respectively. These were well below their expected values for all of the trans conformers and indicated a significant contribution of the more straightened conformers in the nematic phase range.

It is also possible to use the transversal transition dipole moment ($\mu_{\perp}$) to calculate the orientation order of both the short and long molecular axes. In the FTIR spectra of the thioether dimers (CBSCnSCB, CBSCnOCB), we found three bands with a transversal dipole: one at 520 $cm^{-1}$, which could be assigned to the the C-C groups that were out of the benzene plane vibration. This band also involved a thioether bridge and made a significant contribution to the deformation vibration of the sulfur atom *(γCC op CB + δCS)*. Other perpendicular bands were located at the wavenumbers 812 $cm^{-1}$ and were associated with the vibrations of the C-H groups that were out of the benzene plane *(γCH op CB)* and at the 1395 $cm^{-1}$, which were assigned to the in-plane deformation of the C-H groups *(βCH ip CB)*. For the ether dimers (CBSCnOCB, CBOCnOCB), we additionally observed a new maximum at 820 $cm^{-1}$. This complex band, apart from the fact that it was engaged in the deformation vibrations of the C-H group that were out of the benzene plane, also involved the symmetrical stretching vibration of the C-O group *(γCH op CB +* $\nu_S$ *C-O-C)*.

In this case, we had to find two variables, *S* and *D,* thus at least two bands had to be considered in order to solve the problem. In the N phase, however, it was shown that the *D*-order parameter was negligible, and therefore, the *S*-parameter was obtained from the absorbance component of the transversal band (1395 $cm^{-1}$, 812/820 $cm^{-1}$ and 520 $cm^{-1}$) using eq.:

$$S_{ap} = 2 \cdot A_X / A_0 - 2 \qquad (7)$$

## B. Orientational order in the twist-bend nematic phase

At the transition to the $N_{TB}$ phase, the temperature dependence reversed its increasing trend after cooling. This behavior was caused by the gradual tilting of the molecules with respect to the aligning direction. As a result, the order parameter declined from its trend in the N phase, which was extrapolated by power law (5), see Fig. 2. The extrapolation of the *S*-parameter from the N phase ($S_N$ – order parameter of the long axis of the dimer in the N phase) was performed in a temperature range of more than 15 K above the N-$N_{TB}$ transition. As can be seen from the experimental values in the $N_{TB}$ phase, the *S*-parameter was reduced by the $P_2$ Legendre polynomial with respect to the extrapolated parameters (the $S_{TB}$-order parameter of the long axis of the dimer in the twist-bend phase):

$$S_{TB} = S_N \cdot P_2(\cos\theta) \quad (8)$$

where: $\theta$ is the cone angle of the helix (indicates the tilt of the long molecular axis in the $N_{TB}$ phase). Figure 4 shows the sine squared of the tilt angle of the long molecular axis, which was calculated using eq. (8).

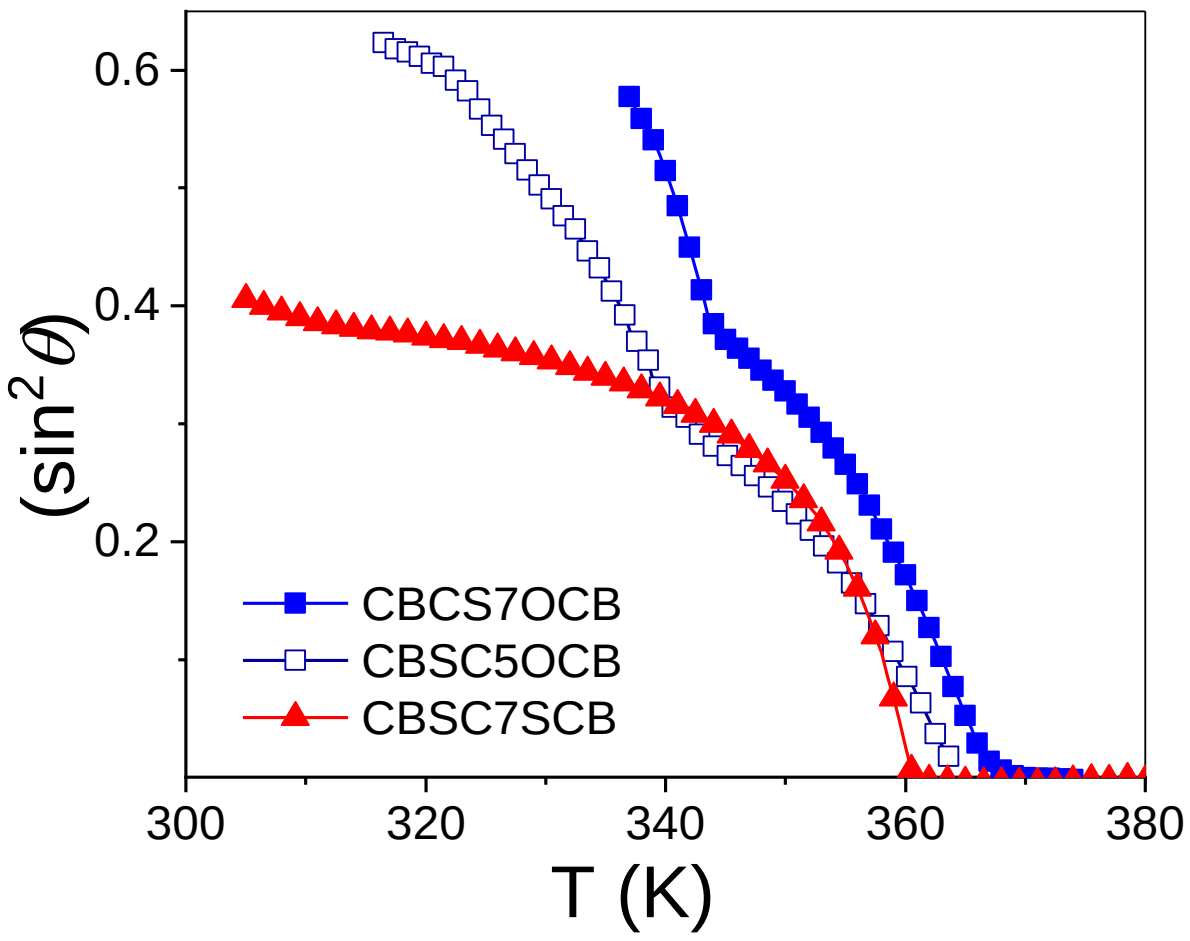

**Figure 4.** The sine squared of the tilt angle of the long molecular axis. Determined from the absorbances of the 1098/1000 cm$^{-1}$ bands. *Symbols:* ■ – CBSC7OCB, □ – CBSC5OCB, ▲ – CBSC7SCB.

### C. Surface anchoring of the dimers

As follows from eq. (2), the average band absorbance, $A_0$, was related to the product of $\left[\frac{d\mu_i}{dQ_i}\right]^2$ [51] and the number density (number of molecules per unit volume). In the range of the N phase, this clearly followed a formula, i.e., the average absorbance increased as the number density of the molecules, which incidentally confirmed that the transition dipole appeared to be constant in this temperature range. In the temperature range below the $N_{TB}$ transition, the behavior was dependent on the substrate of the cell windows and was different for the symmetric and asymmetric dimers.

For the symmetric CBSC7SCB dimer, in the cell composed of polycrystalline ZnSe windows, the average absorbance of the transversal dipole (811 cm$^{-1}$) maintained its trend until the temperature of crystallization. In contrast, the band absorbance that corresponded to the longitudinal transition dipole at the 1600 cm$^{-1}$ (or 2300 cm$^{-1}$, which was assigned to the stretching vibration of the cyano group, $\nu CN$) underwent a significant decrease just after entering the $N_{TB}$ phase, Fig. 5. This behavior clearly indicated that the longitudinal dipoles of the neighboring cyanobiphenyl groups had a tendency to create antiparallel arrangements with respect to each other. The strong dipole of the CN group induced a dipole in the adjacent biphenyl molecule. As a result, we observed the intermolecular van der Waals interactions (more precisely, the London dispersion forces), which led to the stabilization of the $N_{TB}$ phase. We expected that a nano-segregation process might occur along the Z-axis of the system.

This behavior has already been reported for cyanobiphenyl LC, which had an overlapping of both ends of the molecules [49,52]. At the same time, few changes were observed in the perpendicular direction as the absorbance of the transversal dipoles remained unchanged. In the cell composed of polycrystalline ZnSe windows, the average absorbance of the transversal dipole (811 $cm^{-1}$) maintained its trend until the crystallization temperature was reached. However, this behavior changed when we inserted a dimer between the KBr crystalline windows.

Then, in the transition to the $N_{TB}$ phase, we observed an increase in absorbance for the transversal dipole. This increase was even greater for the absorbance of the 1395 $cm^{-1}$ band, see Fig. 5. This difference was probably due to the slightly different azimuthal angle that was created by the corresponding transition dipole moment (1395 or 811 $cm^{-1}$) with the x-z plane of the dimer. It appeared the polar interaction with crystalline KBr influenced the orientation of the transversal dipoles of the dimer and likely induced the biaxiality of the $N_{TB}$ phase [53,54].

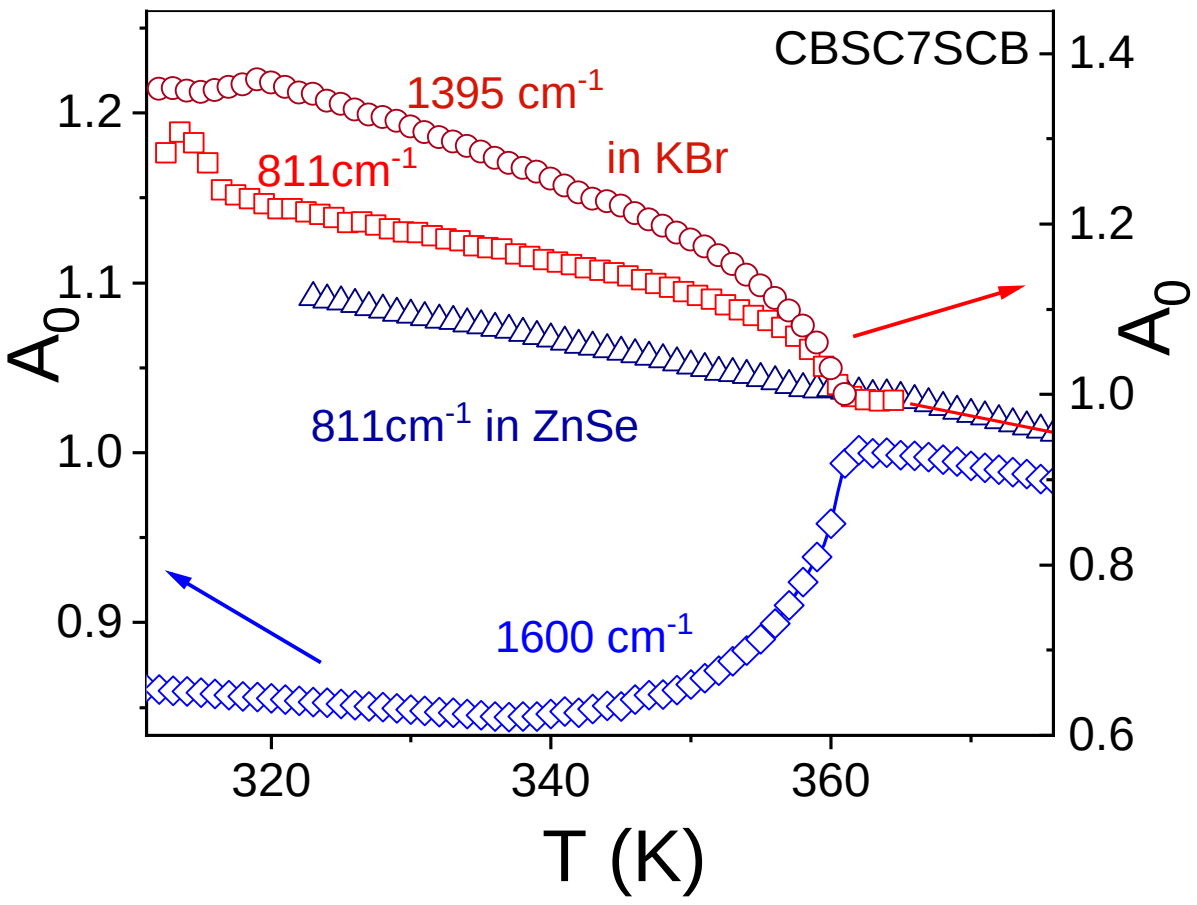

FIG. 5. Comparison of the average absorbance behavior for several bands (811, 1395 and 1600 cm$^{-1}$) of the thioether dimer (CBS7SCB) as a function of temperature depending on the different surface anchoring effects. *KBr substrate*: ○ – *βCH ip CB* at the 1395 cm$^{-1}$ ($\mu_\perp$), □ – *γCH op CB at* 811 cm$^{-1}$ ($\mu_\perp$), *ZnSe substrate*: △ – *γCH op CB at* 811 cm$^{-1}$ ($\mu_\perp$), ◇ – *νCC benzene ring* at the1600 cm$^{-1}$ ($\mu_\parallel$).

For the asymmetric dimers (CBSCnOCB), we could clearly distinguish two regions of the temperature range of the N$_{TB}$ phase: the first extending more than 20 K below the N-N$_{TB}$ transition temperature and the next below ~340 K and approaching almost room temperature. In the first region, the transversal dipole's absorbances maintained their trend (due to an increase in the number density) until the temperature was 340 K, which was similar to the CBSC7SCB dimer in the ZnSe substrate. However, the band absorbance 1600 cm$^{-1}$ (or 2300 cm$^{-1}$) dropped significantly on entering the N$_{TB}$ phase (Figure 6) for the same reason as for the CBSC7SCB dimer. By contrast, in the lower temperature region below 340 K, the lateral dipole changed its behavior with a decreasing temperature and disappeared from the trend that was associated with the density number, which was accompanied by another absorbance drop of the longitudinal dipole.

Such behavior of the absorbance of the transversal dipole may indicate that some new intermolecular interactions had led to the orientation of the bond. It seems that the sulfur group is either involved in the London dispersion forces, π-stacking interactions, or in specific hydrogen bonds with neighboring molecules. It is worth noting that the vibration involving the sulfur linkage (~811cm-1) had an increase in absorbance while that of the oxygen linkage (~820 cm-1) had a decrease in absorbance.

Such a difference in the behavior of the transverse dipoles suggests that the phase symmetry was no longer uniaxial. This was probably induced by the interaction of the transversal dipoles with the cell substrate [55,56]. Moreover, the most probable conformations of the dimers changed and became specific due to the possible rotation around the sulfur bridge.

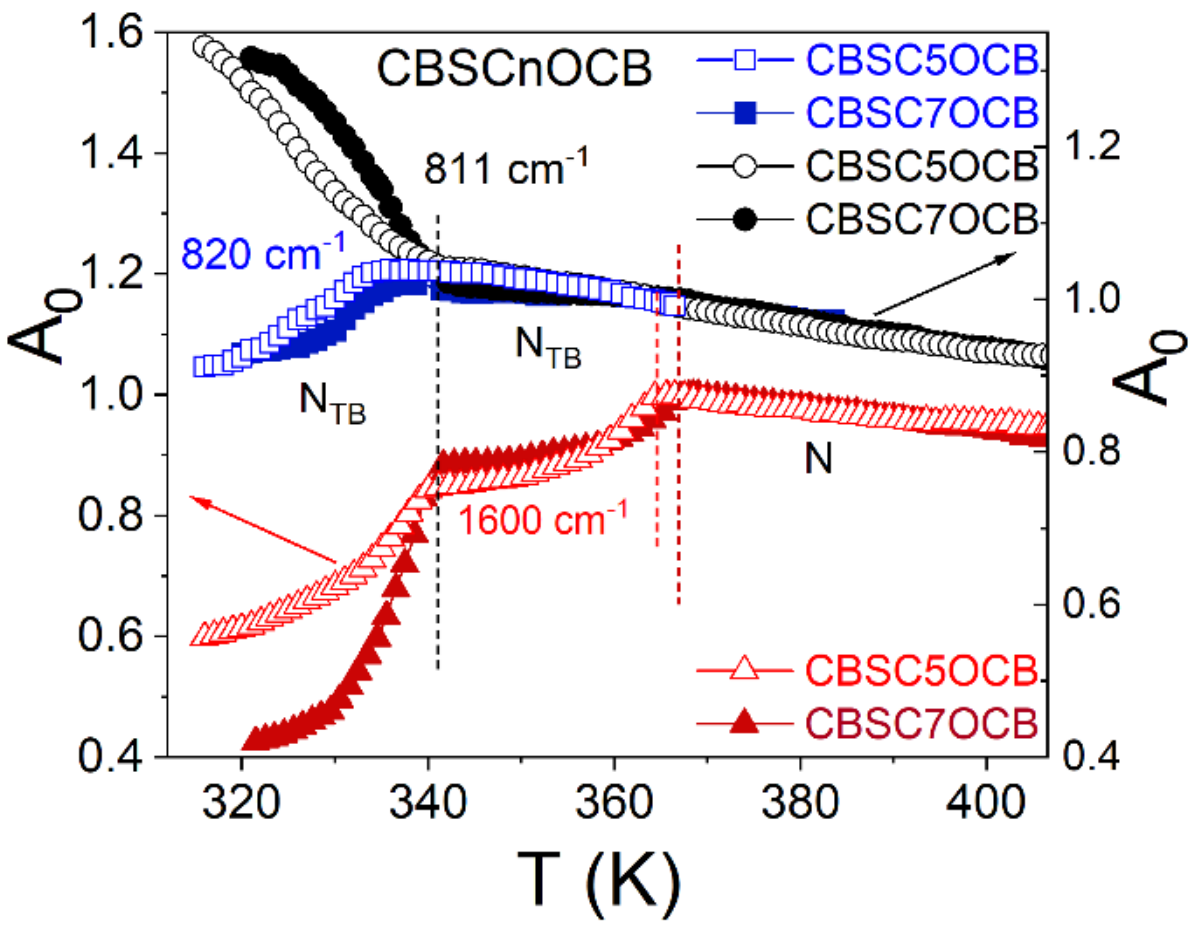

FIG. 6. The average absorbance behavior for several bands (811, 820 and 1600 cm$^{-1}$) of the asymmetric thioether dimers (CBSCnOCB) as a function of temperature. Absorbances of the *transversal* dipole ($\mu_{\perp}$): 811 cm$^{-1}$ (*γCH op CB for the rigid core with a sulfur linkage C-S-C)* ○ – CBSC5OCB, ● – CBSC7OCB; 820 cm$^{-1}$ (*γCH op CB for the rigid core with an oxygen linkage C-O-C)* □ – CBSC5OCB, ■ – CBSC7OCB. Absorbance of the *longitudinal* dipole ($\mu_{\|}$): 1600 cm$^{-1}$ *(νCC benzene ring)* △ – CBSC5OCB, ▲ – CBSC7OCB

Considering the high rotational barrier around the bond connecting the cyanobiphenyl to the alkyl linker, which is much higher in the case of the oxygen bridge than for the sulfur bridge [41], we expected that the oxygen bridge with phenyl group would remain in the **z-y** plane. In contrast, the sulfur bridge might rotate by some/a significant angle with respect to the plane of the phenyl. If we assume that the absorbance changes at 340 K were due to the order rearrangement (phase reorganization), we might consider a phase biaxiality.

Mandle et al. performed a detailed study of bent dimers with different linking groups that yielded a variety of bend angles and linker flexibility. [57]. They showed that the $N_{TB}$ phase could accommodate a range of bend angles, but that the alkyl linker provided the most thermodynamic stability for the phase. Archbold et al. followed up this study by calculating the conformational distributions for several of the reported materials [58]. They found that the thermodynamic stability of the $N_{TB}$ phase results not only from the average bend angle of the molecule but is also dependent on the distribution of the conformers of the $N_{TB}$-forming LC and the flexibility of the spacer. The relationship between the molecular bend and the properties and stability of the $N_{TB}$ phase is still not clear. The LC community has yet to settle on the underlying molecular design rules for $N_{TB}$ materials, which help guide a chemical synthesis beyond meeting the essential requirement of a molecule with a bend. This issue will continue to be a challenge in the future.

### D. Molecular biaxiality.

Figure 7 compares the temperature dependence for the uniaxial $S$-order parameter that was calculated from the vibration with a longitudinal dipole with the apparent parameter that was calculated for the transverse dipole. It is worth noting that the increasing difference between $S_{ap} = 2\cdot A_X/A_0 - 2$ (eq. 7), which was obtained for the transverse transition dipole, did not follow the value that was reported for the longitudinal dipole, $S$. This clearly indicates that the molecular biaxiality, $D$, cannot be further ignored in the $N_{TB}$ phase. Therefore, it is worth determining the biaxiality parameter, $D$, by using the absorbances, $A_X$ and $A_0$, of the transverse dipole and the order parameter for the long molecular axis and the $S$-parameter as follows using eq. (1), $D=2A_X/A_0-2-S$.

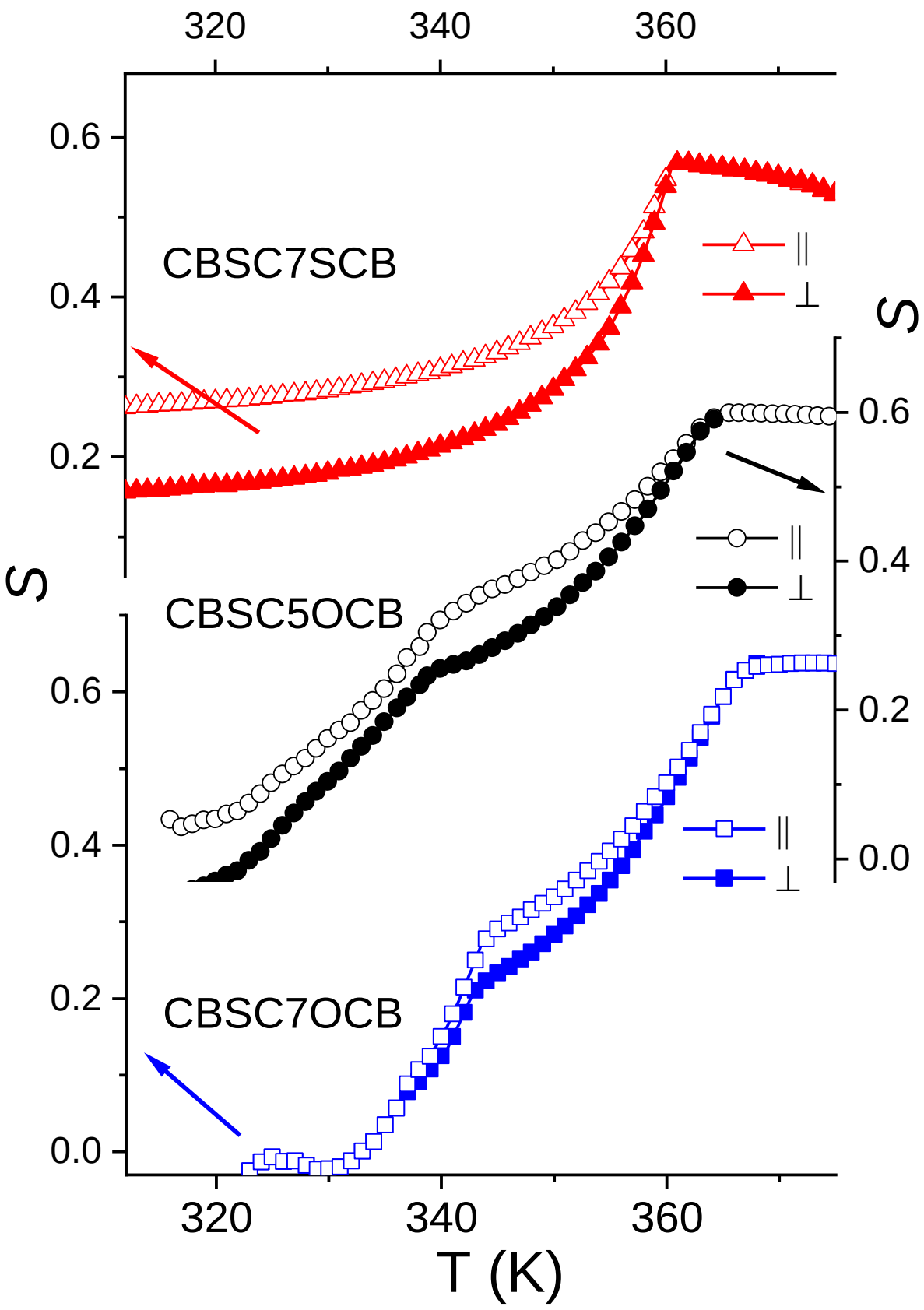

FIG. 7. Comparison of the temperature dependence for the uniaxial $S$-order parameter that was calculated from the vibration with a longitudinal dipole (||) with the apparent parameter $S_{ap}$ that was calculated for the transverse dipole (⊥). *Open*: $S$ for the longitudinal dipole (||) calculated from the mode at 1098cm$^{-1}$ and 1000 cm$^{-1}$ for the symmetric (CBSC7SCB) and asymmetric dimers (CBSC5OCB, CBC7OCB), respectively. *Solid*: transverse dipole (⊥) calculated from the mode at 811 cm$^{-1}$ for all of the dimers.

Figure 8 shows the temperature dependence of the calculated molecular biaxiality for the asymmetric dimers (CBSC5SOB, CBSC7OCB) and the symmetric CBSC7SCB dimer in the temperature range of the $N_{TB}$ phase. The results were calculated for two bands: 1098 cm$^{-1}$, which corresponds to the longitudinal dipole, and 811 cm$^{-1}$, which describes the transverse dipole for the CBSCnSOB dimers. In the case of the CBSC7SCB dimer, we selected the band at the 1395 cm$^{-1}$ as the transverse dipole.

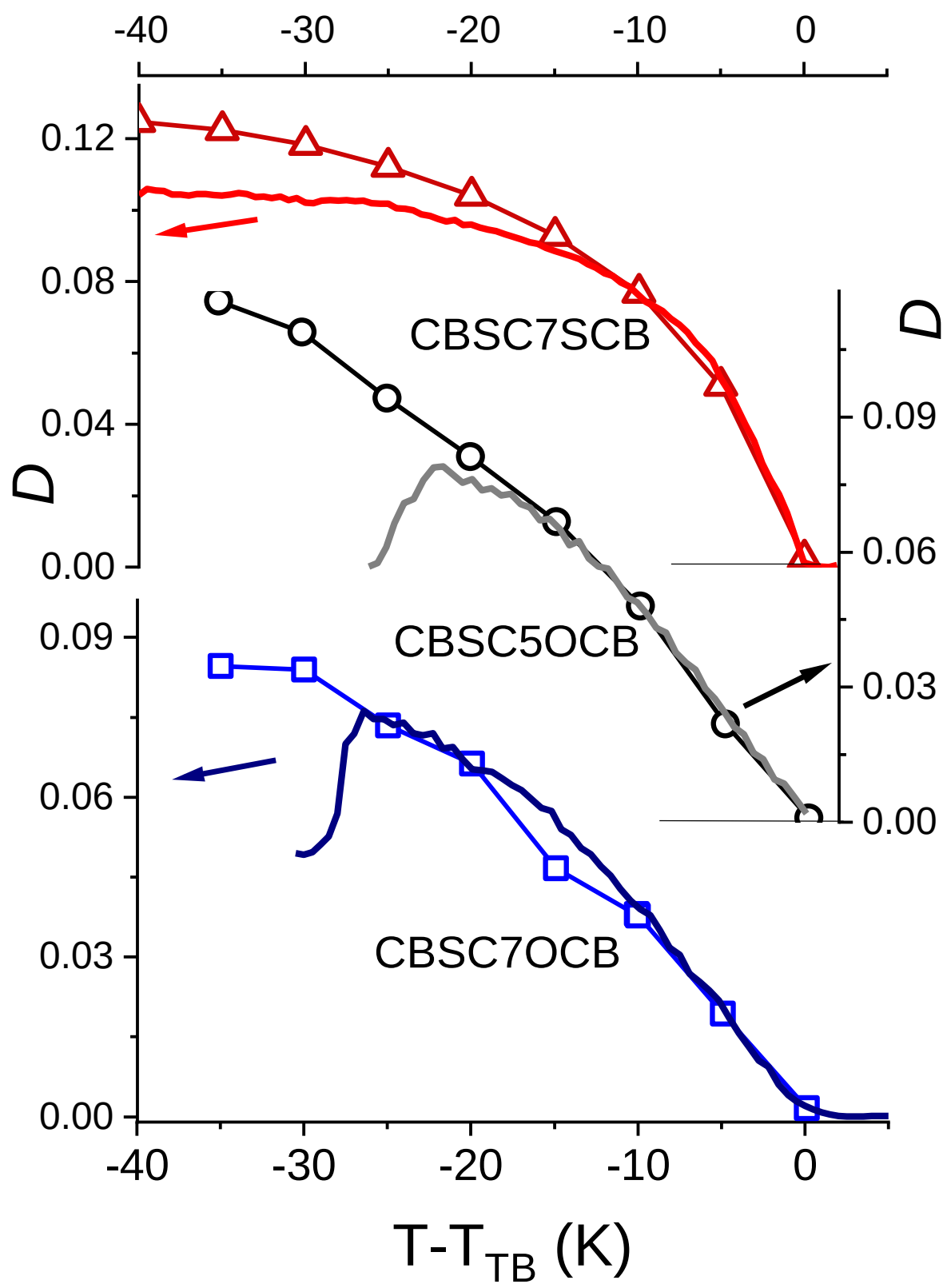

FIG. 8. Molecular biaxiality parameter *(D). Solid lines: navy* – CBSC7OCB, *black* – CBSC5OCB, *red* – CBSC7SCB. The local bending, <$\sin^2(\beta_2)$> calculated for the dimers: □ – CBSC7OCB, O – CBSC5OCB, △ – CBSC7SCB.

By the definition of the pitch, we have $q \equiv d\alpha/dz$, and from the geometry of the helix, we can connect $q$ [41] with the bend magnitude of the helix, $B=q\cos\theta\sin\theta$ [44], where $\alpha$ is the azimuthal rotation of the director per half the molecule length, $d_2$ (see in Figure 9). Then, the molecular bend can also be associated with the bend magnitude of the helix:

$$\sin\beta = \mathrm{tg}(\alpha/2)\sin\theta \cong 0.5 \cdot d_2 q \cos\theta \sin\theta \tag{9}$$

It was already shown [45] that the director bending, $\sin\beta/(d_2\cos\theta)$, is linearly dependent on $\sin\theta$. In a molecular system, however, the bending vector has to be considered statistically, and therefore its actual size, $<\sin^2\beta>^{0.5}$, is the result of transforming the bend of the molecule, $\sin\beta$, into its statistically averaged value, i.e., the view of the molecule that is perceived by the system:

$$\langle \sin^2\beta \rangle \cong D\sin^2\beta \tag{10}$$

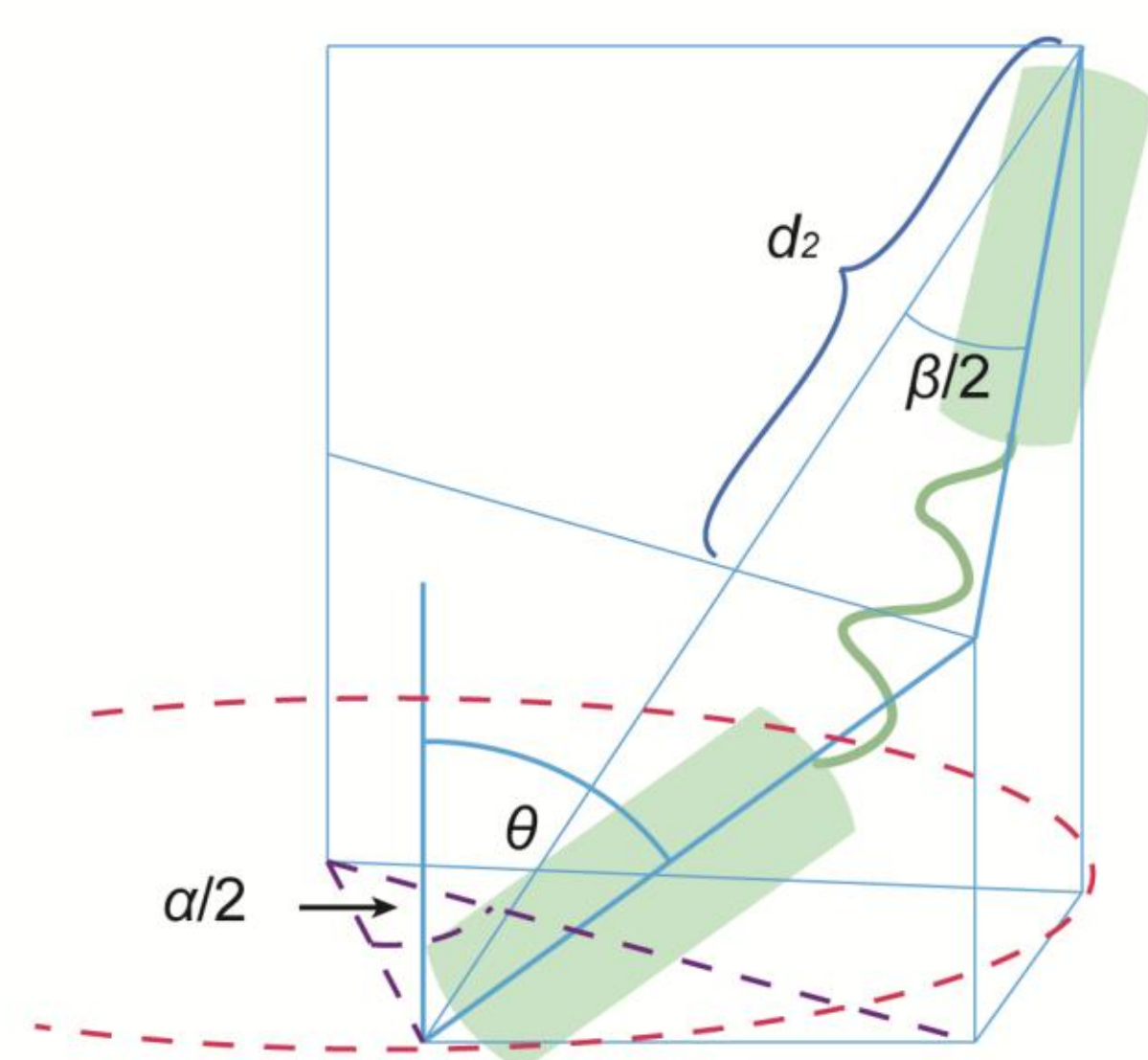

**FiG. 9.** Geometrical model of the helical structure composed of dimer molecules. The model shows the azimuthal rotation, $\alpha/2$, of the director per half the molecule length, $d_2$. Here: $\theta$ is the molecule tilt and is the $\beta$ bend of the molecule.

As a result, the molecular biaxiality parameter, $D$, due to eq (9) and eq (10) is accordingly dependent on $\sin^2\theta$.

$$D \cong \langle \sin^2 \beta \rangle / \left( {d_2}^2 \sin^2 \beta \right) \cong C \sin^2 \theta \tag{11}$$

where: $C=(d_2 \sin\beta)^{-2}$ is the slope of the "$D$ vs. $\sin^2\theta$". As was previously shown for terphenyl dimers [33], the $D$-parameter can be associated with the square of the tilt angle, Fig. 10.

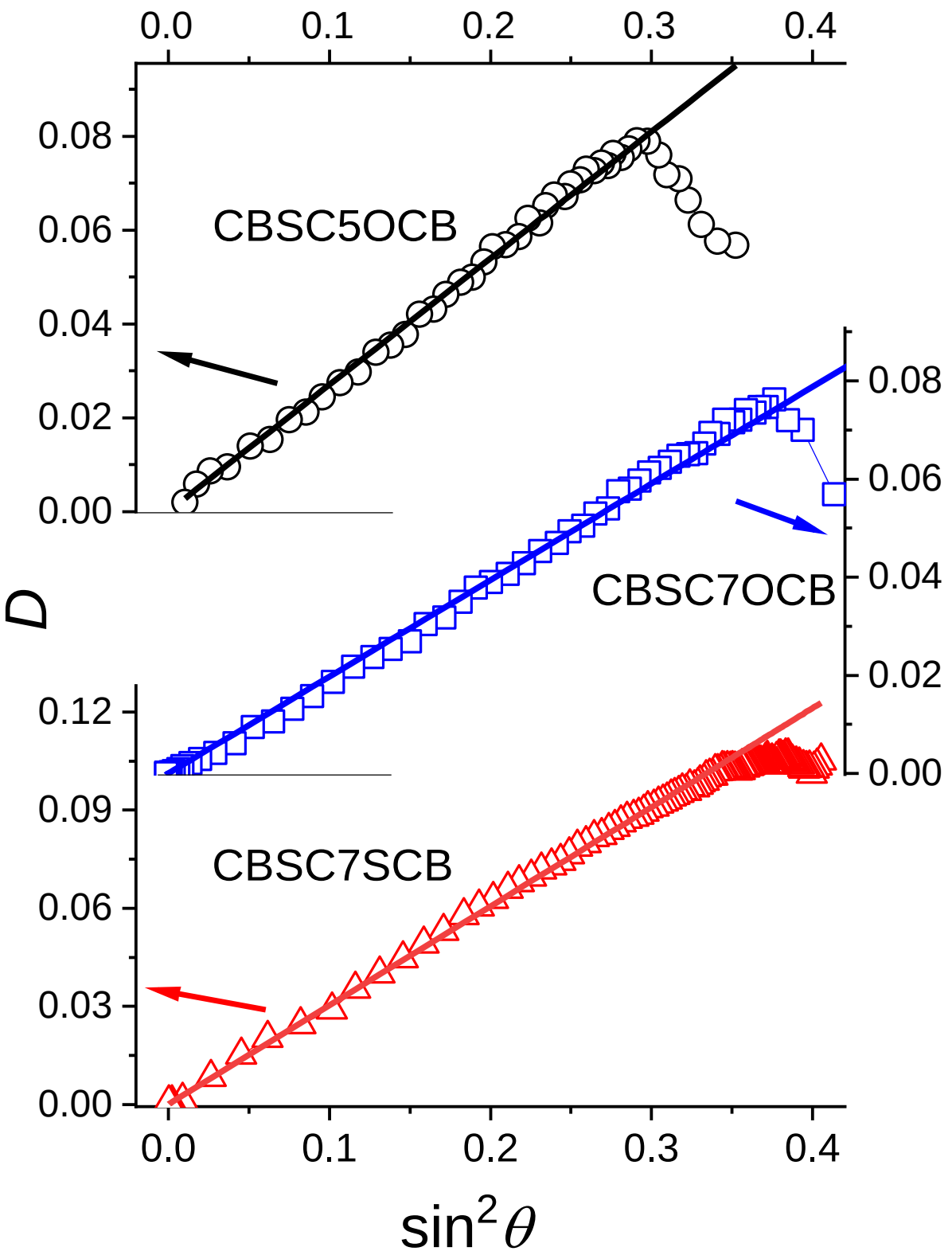

**FIG. 10.** Molecular biaxiality parameter, $D$ vs. $\sin^2\theta_t$, for the thioether dimers.

□ – CBSC7OCB, O – CBSC5OCB, △ – CBSC7SCB and the solid lines are the linear fits to the data (0.0 origin).

We compared the behavior of several series of LC dimers. The chemical nature of the linking groups between the mesogenic units and the spacer was systematically varied in order to include methylene, ether and thioether units. For the dimers containing an odd number of atoms in the spacer, the ether linked is the most linear and the thioether linked is the most bent. This increasing molecular curvature decreases the N-isotropic transition temperature and affects the birefringence of the nematic phase. Theoretical studies of the rigid V-shaped molecules predict that the $N_{TB}$ phase formation and its stability depends on molecular curvature [59,60] Experimentally, the N-$N_{TB}$ transition temperature is the highest for the dimers containing a methylene spacer and is similar for the corresponding ether and thioether materials. However, the thioether dimers are more bent than their methylene counterparts. It appears that the optimal molecular curvature for the formation of the $N_{TB}$ phase is around 120°.

Surprisingly, despite the different molecular structures, the helical pitch length, which is far from the N-$N_{TB}$ transition, corresponded to four longitudinal molecular distances in all of the studied materials. On approaching the transition to the N phase, the pitch began to unwind and the number of molecules per turn increased critically with the N-$N_{TB}$ transition being weakly of the first order. There is much work yet to be done, both experimental and theoretical, in order to understand these fascinating systems.

## IV. CONCLUSIONS

Concerning the molecule shape – the dimers that more bent were found to be less compatible for creating the nematic order. Thus, the opening angle must have a significant impact on the orientational order parameter. An increase of bend angle relatively reduced the orientational order. Namely, the temperature dependence became weaker as the temperature dropped. The molecular biaxiality parameter was found to be negligible in the N phase and then began to increase upon entering the $N_{TB}$ phase. The local director deformation was found to be well correlated with the molecular biaxiality parameter *D*. Changes in the mean absorbance in the $N_{TB}$ phase were observed as a result of the intermolecular interactions: a decrease for the longitudinal dipole at the N-$N_{TB}$ transition, which emphasized the antiparallel axial interaction of the dipoles, while the absorbance of the transverse dipoles remained unchanged up to 340 K when the latter became parallelly correlated.

The main problem, which had some limitations in the results, was the great deal of difficulty in obtaining a homeotropic alignment for the cyanobiphenyl dimers. Measurements of all three components of the absorbance ($A_X$, $A_Y$, $A_Z$) requires at least two types of alignment (planar & homeotropic). In the case of the dimers with a cyanobiphenyl core, an attempt to determine the homeotropic order using the standard methods (a suitable surfactant or a strong magnetic field) led to a hairpin conformation of the molecules more than a bent conformation. In this paper, we determined one perpendicular and one parallel absorbance component. To calculate the average $A_0$, we assumed that the perpendicular component $A_X$ was equal to $A_Y$ ($A_\perp = 2*A_X$). This approach led to the assumption that the observed twist-bend phase was uniaxial. In the case of our considerations on the surface-induced biaxiality below 340 K, it was necessary to measure the second perpendicular component in order to unequivocally calculate the biaxial phase order parameters (*C, P*).

This work will be the basis for developing a model that, based on the molecular parameters that are obtained from polarized FTIR spectra measurements and a DFT structure simulation, will enable the essential properties of the $N_{TB}$ phase such as cone angle, director bending and indirectly helical pitch to be predicted. The development of such a model will enable the theoretical estimation of the helix pitch parameter, which is extremely important when such a measurement is impossible. The most crucial issue, especially in photonic applications, is spatial periodicity.

AUTHOR INFORMATION

**Corresponding Authors:**

* E-mail: katarzyna.merkel@us.edu.pl (K.M)

**E-mails co-authors:**

Antoni Kocot: antoni.kocot@us.edu.pl

Barbara Loska: barbara.loska@us.edu.pl

Yuki Arakawa: arakawa.yuki.xl@tut.jp

**ORCID:**

Antoni Kocot: 0000-0002-9205-449X

Barbara Loska: 0000-0002-0756-5018

Yuki Arakawa: 0000-0002-8944-602X

Katarzyna Merkel: 0000-0002-1853-0996

**Author Contributions:**

‡ K.M and A.K contributed equally to this work.

Conceptualization: K.M, A.K
Synthesis: Y.A
Methodology: K.M, A.K
Investigation: K.M, B.L, A.K
Writing—original draft: A.K, K.M
Writing—review & editing: K.M

**Funding Sources:**

National Science Centre, Poland for Grant No. 2018/31/B/ST3/03609

National Science Centre, Poland for Grant No. 2020/39/O/ST5/03460

**Competing interests:**

Authors declare that they have no competing interests.

ACKNOWLEDGMENT

Authors K.M. & A.K. thank the National Science Centre for funding through the Grant No. 2018/31/B/ST3/03609. B.L. thanks the National Science Centre for funding through the project No. 2020/39/O/ST5/03460.

All DFT calculations were carried out with the Gaussian09 program using the PL-Grid Infrastructure on the ZEUS and Prometheus cluster.